\documentclass[sigconf]{acmart}
\renewcommand\footnotetextcopyrightpermission[1]{} 
\usepackage{tikz}
\usepackage{amsmath}
\usepackage{tabularx}
\newcolumntype{Y}{>{\centering\arraybackslash}X}

\newcolumntype{Z}{>{\hsize=1.2\hsize}X}
\newcolumntype{Q}{>{\hsize=.8\hsize}X}
\newcolumntype{V}{>{\hsize=.15\hsize}X}

\usepackage{caption}
\usepackage{subcaption}
\usepackage{stfloats}
\usepackage{multirow}
\usepackage{multicol}
\usepackage{makecell}
\usepackage{calc}  
\usepackage{enumitem} 
\usepackage{nicematrix,tikz}
\usepackage{xurl}

\usepackage{filecontents}

\AtBeginDocument{%
  \providecommand\BibTeX{{%
    \normalfont B\kern-0.5em{\scshape i\kern-0.25em b}\kern-0.8em\TeX}}}

\setcopyright{acmcopyright}
\copyrightyear{2018}
\acmYear{2018}
\acmDOI{XXXXXXX.XXXXXXX}

\acmConference[Conference acronym 'XX]{Make sure to enter the correct
  conference title from your rights confirmation emai}{June 03--05,
  2018}{Woodstock, NY}
\acmPrice{15.00}
\acmISBN{978-1-4503-XXXX-X/18/06}

\begin{document}

\title{Gone Quishing: A Field Study of Phishing with Malicious QR Codes}

\author{Filipo Sharevski}
\affiliation{%
  \institution{DePaul University}
  \streetaddress{243 S Wabash Ave}
  \city{Chicago}
  \state{IL}
  \postcode{60604}
    \country{United States}
}
\email{fsharevs@cdm.depaul.edu}

\author{Amy Devine}
\affiliation{%
  \institution{DePaul University}
  \streetaddress{243 S Wabash Ave}
  \city{Chicago}
  \state{IL}
  \postcode{60604}
    \country{United States}
}
\email{adevine@depaul.edu}

\author{Emma Pieroni}
\affiliation{%
  \institution{DePaul University}
  \streetaddress{243 S Wabash Ave}
  \city{Chicago}
  \state{IL}
  \postcode{60604}
    \country{United States}
}
\email{epieroni@depaul.edu}

\author{Peter Jachim}
\affiliation{%
  \institution{DePaul University}
  \streetaddress{243 S Wabash Ave}
  \city{Chicago}
  \state{IL}
  \postcode{60604}
    \country{United States}
}
\email{pjachim@depaul.edu}

\renewcommand{\shortauthors}{Authors}

\begin{abstract}
The COVID-19 pandemic enabled \textit{quishing}, or phishing with malicious QR codes, as they became a convenient go-between for sharing URLs, including malicious ones. To explore the quishing phenomenon, we conducted a 173-participant study where we used a COVID-19 digital passport sign-up trial with a malicious QR code as a pretext. We found that 67 \% of the participants were happy to sign-up with their Google or Facebook credentials, 18.5\% to create a new account, and only 14.5\% to skip on the sign-up. Convenience was the single most cited factor for the willingness to yield participants' credentials. Reluctance of linking personal accounts with new services was the reason for creating a new account or skipping the registration. We also developed a \textit{Quishing Awareness Scale} (QAS) and found a significant relationship between participants' QR code behavior and their sign-up choices: the ones choosing to sign-up with Facebook scored the lowest while the one choosing to skip the highest on average. We used our results to propose quishing awareness training guidelines and develop and test usable security indicators for warning users about the threat of quishing. 
\end{abstract}

\ccsdesc[500]{Security and privacy~Social aspects of security and privacy}
\ccsdesc[500]{Security and privacy~Usability in security and privacy}

\keywords{phishing, QR codes, quishing, usable security}


\maketitle

\section{Introduction}
Disruptive in every aspect, the COVID-19 pandemic forced a practical reorganization of our rather comfortable lifestyles. The ergonomic aspect in particular took the hardest hit, outside the devastating public health toll: Mask and distance mandates, virtual work environments, contact-free exchange of goods, and touch-free information transfer \cite{CDC-Guidance}. Adopting alternative designs for otherwise intuitive everyday interactions was inevitable to prevent the COVID-19 virus from rapidly spreading and we collectively embraced the task at hand \cite{CDC-Resume}. Instead of paper, we got to scan a Quick Response (QR) code with our smartphones to read a restaurant menu or enter a venue, for example. QR codes, initially developed for high-speed component scanning, are not an entirely new design per se, but were far from being widespread before the requirement for touch-free interaction \cite{Tai-Wei}. 

QR codes embed a simple piece of information, either a (i) unique identifier that enables external scanning and verification against a back-end database (e.g. barcodes) or (ii) Universal Resource Locators (URLs) that, once decoded through users' smartphone camera, enable direct access to websites (e.g. restaurant menus) without the need to type the link themselves \cite{Mayrhofer}. The simplicity of the information piece itself made the QR codes the prefect candidate for manipulation. But beside impersonation or duplication, malicious actors were yet to entertain the full potential of manipulating the QR codes to their advantage  \cite{Yijie}. However, the more applications adopted QR codes, the more creative attacks exploiting a wide range of victims surfaced \cite{Kharraz}.      

This is especially true for phishers \cite{Krombholz}. Controlling malicious URLs (e.g. links to impersonated login websites), phishers are equally keen to adopt new alternative ``designs'' in luring victims to yield their credentials or download malware \cite{Kieseberg}. Phishers started simple by luring victims to malicious URLs through unsuspected QR codes \cite{Focardi, Mavroeidis}. They moved next to embed a malicious URL in a QR code that was itself embedded in another QR code knowing that the inner one is scanned first before the outer, legitimate one \cite{Dabrowski}. Phishers even became prolific in sophisticated alteration of the codes pixels of unsuspected QR codes to embed the malicious URL without the need for physical tampering \cite{zhou}. 

The practice of ``\textit{quishing}'' or phishing with malicious QR codes (i.e. QR codes that embed a malicious URL) affected a small number of victims and thus never received much of an attention \cite{Chouinard}. QR codes' lack of widespread adoption certainly was the main limiting factor in the pre-COVID-19 world. But quishing popularized when QR codes became the essential go-between for information sharing and the collective effort was concentrated on containing a \textit{real} virus, not a ``computer'' one. Providing a fast, touch-free URL transfer, QR codes' convenience primed the users to accomplish a task (e.g. read a restaurant menu) without paying much attention to any possible threats of phishing exploitation \cite{Bai}. 

It comes to little surprise, then, that the quishing attacks not just rose in numbers, but diversified the tactics and types of pretexts employed \cite{fbi_qr}. For example, malicious QR codes were found on parking meters throughout San Antonio, directing victims to a fraudulent website to submit payment to a fraudulent vendor \cite{Barr}. Quishers attached same-day generated malicious QR codes to emails to evade URL detection (unlikely to be updated to block QR code images) in a phishing campaign attempting to collect Microsoft credentials \cite{Chouinard}. And quishers elegantly included a backdoor in an open-source QR code generator that contains malware which they could use to remotely execute code on a compromised machine or install and access a remote shell \cite{Murphy}. 

The proliferation of quishing certainly warrants a response, not just by issuing user advisories \cite{fbi_qr}, but also exploring other pretexts and tactics, gauging users' susceptibility, and raising users' awareness of yet another phishing avenue. To these objectives, we ``went quishing'' in a field study with 173 participants. The pretext we created was a sign-up for a COVID-19 digital passport trial aiming to replace the paper vaccination cards issued by the Centers of Disease Control (CDC) \cite{CDC-Card}. We incorporated a malicious QR code within a legitimate-looking CDC poster borrowed from their Toolkit for General Public communication \cite{CDC-Public}. We observed the sign-up choice for each participant (Facebook or Google credentials, creating a new account, or skipping a sign-up) and collected information about participants' practice of using QR codes. 

Our observations show that 67 \% of the participants opted to use either their Facebook or Google credentials, 18.5\% to create a new account, and only 14.5\% to skip on the sign-up. The analysis of the choice justification by each participant provides further evidence that users would yield their credentials as long as that achieves a minimum compliance and avoids extra steps, as is the case with any other form of phishing \cite{Wash}. We observed a positive effect of online safety cognizance where the participants opted for the middle ground and created an account or skipped the sign-up entirely. In the former case, the justification provided was to keep their Facebook/Google accounts safe from a ``\textit{seemingly unfamiliar website}.'' In the latter case, the justification was to \textit{``avoid identification and sharing of private medical information.''} 

We developed a \textit{Quishing Awareness Scale} (QAS) following the approach for measuring proactive phishing awareness in \cite{Egelman}. We found a significant relationship between participants' QAS score and their sign-up choices: the participants who opted for the Facebook route scored the lowest and the participants who opted to skip the sign-up scored the highest on the QAS scale. The results of our study motivated us to create actionable quishing awareness training guidelines incorporating the wealth of previous phishing awareness knowledge and practice \cite{Wash, Franz}. We also utilized various usable security recommendations to develop and test practical anti-quishing security indicators in a follow-up study \cite{Schechter, Petelka, Wogalter}.






Following this introduction, we outline the past and present approach of quishing in Section \ref{section:2}. We then elaborate on our pretext, quishing tactics, and measurements in Section \ref{section:3}. Section \ref{section:4} provides the results of our study. Section \ref{section:5} outlines our anti-phishing training guidelines and the usability tests we conducted to evaluate quishing indicators we proposed. The overall findings are discussed in Section \ref{section:6} and Section \ref{section:7} wraps the paper.

\section{Background Research} \label{section:2}
\subsection{How Users Fall for Phish}
Phishing, in the most common example, involves an attacker sending an email asking the user to click on a URL and enter in their credentials on the resulting website \cite{Wash}. The resulting website does mimic the appearance of a website from a trusted organization (e.g. a bank or Gmail), but it does so on the surface. In the background, the resulting website is associated with a malicious URL and is controlled by the attacker. The user has no way of knowing, unless carefully inspecting the URL and the website's layout \cite{Nicholson}, that the entered credentials end up in the attackers hands instead \cite{Redmiles2019}. 

Phishing works time and again for several reasons. The ``asking'' part in the email incorporates a persuasive pretext sufficiently potent to incite an action by the user. The attacker knows well that there are fundamental vulnerabilities of human cognition  - or 'shortcuts' - that determine decisions on the basis of previous experiences, biases, or beliefs \cite{Cialdini}. So when the email is ``signed'' and comes from a bank or Gmail, the user feels compelled to comply with these \textit{authorities} \cite{iost2020}. Even more, if the email says that the user must ``update the account within 24 hours or loose access,'' then the user feels the \textit{urgency} to do so \cite{Lin}. Sometimes, the asking comes with a small reward (``sign-up with your Gmail account, receive an Amazon gift card'') making the user feel obliged to \textit{reciprocate} \cite{Blythe}. Sometimes, users are provided a \textit{social proof} that other users, some of whom they \textit{like} or share similarities with, already took the action asked \cite{Ferreira}. 

Despite being persuaded to click on the link, users still have an option to avoid being phished. The resulting websites sometimes present typographical errors and visual discrepancies related to the typeface, design layout, or logos \cite{Fernandez}. So users, if paying attention, could spot misspellings, grammatical errors, inconsistency between mobile and desktop versions, and perhaps deprecated logos. But that's not always the case and these discrepancies remain unnoticed by most of the phishing victims \cite{Alsharnouby}. Even if the resulting website appearance could be indistinguishably mimicked, attackers must deviate the malicious URL from the legitimate one to redirect the user \cite{Thompson}. Spotting a malicious URL remains a hard problem for humans \cite{Reinheimer}, but modern email clients and browsers are better in detecting the malicious URLs, so they employ preemptive indicators of potential phishing to warn the users \cite{Reeder}. Users, unfortunately, often ignore these indicators because they are confusing, interfere with the primary task of entering credentials, or users are not technically proficient to understand them \cite{Porter-Felt}.  

\subsection{How Users Fall for Quish}
In quishing, or phishing with QR codes, the attacker does not have the advantage of using an email. Here the malicious URL has been either embedded in the QR code itself \cite{Mavroeidis, Dabrowski} or the QR code layout manipulated to redirect the user to override the URL of the trusted organization \cite{Kieseberg, zhou}. Assuming this is successful, the attacker must adapt the persuasive pretext to fit with the natural usage of QR codes i.e. all 'shortcuts' might not work. Looking \textit{authoritative} by adding QR codes in an email supposedly coming from a bank or Gmail did yield some success \cite{Chouinard}, however email systems could filter out any message with a QR code. The next thing is to try to assume authority of a convenience service like in the case of parking meters \cite{Barr} or bus stop schedules \cite{Mavroeidis}, but one could simply use a smartphone application instead of a QR code.      

Assigning \textit{urgency} (\textit{scarcity} in general) for the purpose of scanning a QR code could work given that users that fell for quish so far did it so mostly out of curiosity \cite{Vidas}. In this case, coupling with \textit{reciprocation} could work well too, because it creates a tangible proposition for a user to scan a QR code in the first place, e.g. participate in a study. The \textit{social proof} is naturally helpful in such pretext, provided the QR codes are placed where usual study recruitment takes place - college campuses, for example - because others also participate too. The topic for the study could be something that naturally fits what a user in such a setting might like: self-determination, social networking, or campus life. 

Another conducive factor to quishing, at least for now, is that humans do not get much of a help in spotting malicious QR codes. Incorporating security primitives in QR codes could help, but creates an overburdening computational delay (e.g. users start aborting the scanning) \cite{Mavroeidis, Focardi}. A trusted organization could create a QR code with distinctive properties, e.g. logos or a complex color scheme \cite{Krombholz}, but it is increasingly trivial for attackers to duplicate and impersonate such codes. The smartphone's QR scanning function and browsers could incorporate usable security indicators, but nothing says users would not ignore these too \cite{Carroll}. Quishing advisories, as the one from the Federal Bureau of Investigation (FBI) \cite{fbi_qr}, do attempt to raise awareness but are far from a comprehensive approach comparable to the awareness work done for phishing \cite{Reinheimer, Franz}.

\section{Malicious QR Codes} \label{section:3}

What is still largely unknown is how users could fall for quish if the pretext is outside of a college campus setting or places like bus stops with limited number of people. The established trusted norm of using QR codes during COVID-19 provides an ideal opportunity to do such tests since it drastically expanded the number of users as potential participants meeting the ``inclusion criteria'' for quishing studies. Now almost everyone has used a QR code at least once already and chances are high we will keep doing so in the foreseeable future \cite{CDC-Resume}. So a quishing study in these settings must not just explore the workings of quishing on a larger scale, but produce actionable deliverables of immediate anti-quishing help. To these points, we design a usability study pertaining to answer the following research questions: 

\begin{description}
\item \textbf{RQ1}: What susceptibility factors contribute in a plausible quishing campaign for harvesting credentials targeting a general user population? 

\item \textbf{RQ2}: How one's quishing awareness factors in their willingness to either yield their social media and email credentials, provide new credentials, or avoid providing any credentials at all?  

\item \textbf{RQ3}: How the results from a quishing field study could be incorporated with past knowledge in creating training for users not to fall for quish? 

\item \textbf{RQ4}: How the results from a quishing field study could inform the design of security indicators to help users spot malicious QR codes?

\end{description}

\subsection{Quishing Pretext}
The pretext in our study employed several elements targeting the users' shortcuts regularly exploited in both phishing and quishing so far. To appear \textit{authoritative}, the pretext invited participants to sign up for a COVID-19 digital passport trial aiming to replace the paper vaccination cards issued by the Centers of Disease Control (CDC) \cite{CDC-Card}. The attention to CDC during the COVID-19 pandemic ensured participants have heard of it and it was successfully used as an authority in phishing already \cite{CDC-Phishing}. Unlike other countries' health authorities, CDC has not yet issued a digital vaccination passport despite the convenience of such tool for proof-of-vaccination in businesses or travel \cite{Dye}. Efforts to do so were signaled by the US health authorities, but nothing yet has materialized up and during our study window, so it was fairly reasonable for participants to believe a trial for vaccination passports would take place (trials during the initial COVID-19 vaccine development received detailed media attention, for example \cite{Martin}). 

\begin{figure}[!h]
  \centering
  \includegraphics[width=\linewidth]{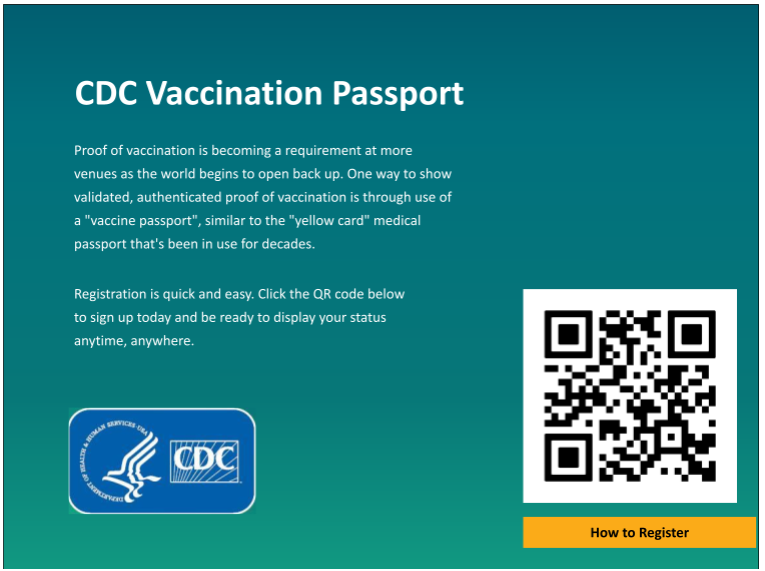}
  \caption{Quishing Pretext Flyer}
  \label{fig:pretext-flyer}
\end{figure}

Using the CDC's media communication toolkit and \cite{CDC-Public}, we developed a flyer for the COVID-19 vaccination passport trial shown in Figure \ref{fig:pretext-flyer}. We paid attention to mimic the appearance of the regular CDC communication on social media assuming most of the users would have already been exposed to posts with very similar design. The flyer also enabled us to overcome the absence of a traditional email communication and include a verbose persuasive text. We used an authoritative yet optimistic premise for the need of a digital vaccination passport and provided an analogy with the traditional ``yellow card'' passports  \cite{Gigerenzer}. We employed \textit{scarcity} by pointing out the ``proof-of-vaccination requirement'' as a  as well as implicit \textit{reciprocation} and \textit{social proof} since the ``sign-up is quick and easy; you can use the digital passport anytime, anywhere.'' Participants, we assumed, have already felt the analogous social proof with the paper vaccination cards and would reciprocate with participating in the trial - by scanning the QR code in the flyer - for the convenience of having a proof-of-vaccination on their smartphones instead.

\subsection{Quishing Infrastructure}
\begin{figure}[!h]
  \centering
  \includegraphics[width=0.52\linewidth]{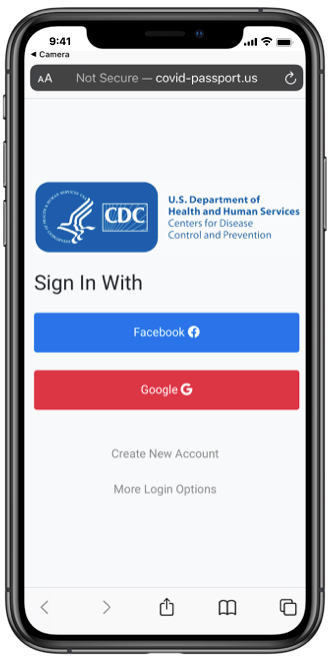}
  \caption{Quishing Landing Page}
  \label{fig:landing}
\end{figure}

The QR code from the flyer directed the users to \url{http://covid-passport.us/qr}, a URL we established just for this project. The landing page of the resulting website is shown in Figure \ref{fig:landing}. We didn't choose to use a secure URL because we wanted to create a realistic quishing scenario where the participants have the opportunity to use the security indicators employed by standard smartphone browsers. For example, Figure \ref{fig:landing} shows the \texttt{Not Secure} text indicator preceding the URL in the browser bar of the Safari browser on an iPhone warning the user of a potentially insecure website ahead (Chrome shows a white exclamation mark in a black triangle and Firefox uses a red backslash over a padlock icon). 

The landing page retains the CDC logo we used in the flier and employs a minimal design centered around the options for sign-up. We ran an experimental pilot study to determine the most popular ways of signing up that confirmed Facebook and Google as the preferred Single Sign On (SSO) providers \cite{Bauer}. We mimicked the respective SSO buttons to lead users to Facebook and Google login options. We also included the option for the participants to create a new account or to skip using any credentials. To avoid registration, they had to first click on the ``More Login Options'' link which showed the a ``Skip'' button. We deliberately introduced an extra step to observe if the participants would take it to stay secure.  

The Institutional Review Board (IRB) did not allow us to collect participants' actual credentials. When the participants clicked the SSO and were about to land on Facebook and Google impersonated login options, we redirected them to an individual survey link where they were asked open-ended questions to rationalize the choice of login, elaborate on their individual use of QR codes, and answer quishing awareness-related questions (in a randomized order for each participant). Both reputation and attention checks were included to prevent machine-generated inputs and poor responses. The survey was anonymous and allowed users to skip any question they were uncomfortable answering.

We did obtain a non-full disclosure approval from the IRB which required collecting a consent before the exposure to the quishing poster. Our inclusion criteria stipulated participants to be from the US, 18 years old, and familiar with CDC vaccination cards. Each participant, recruited through Amazon Mechanical Turk, was initially directed to a separate consent page that contained the poster with the QR code for the trial. Each of the branching surveys included a debriefing statement at the end explaining the deception used in the quishing study. Participants had the option to ask their data to be removed (no one did so). Participants were compensated by the standard rate for a 20 minute study participation, which we advertised as a ``COVID-19 Digital Vaccination Passport Trial.'' 


\subsection{Quishing Awareness Scale}
The stipulation for no direct collection of participants' credentials helped us avoid situations where the participants would drop the survey feeling being tricked. Retaining participation was useful for us to collect data and measure participants' underlying psychological constructs when interacting with QR codes. Since quishing is a variant of phishing, we looked at adapting existing scales that measure end-user behaviors when dealing with malicious URLs. One such scale is the \textit{Security Behavior Intentions Scale (SeBIS)} \cite{Egelman}. 

The SeBIS scale consists of 16 items that measure 4 underlying constructs: device securement, password generation, proactive awareness, and updating behaviors. Interacting with QR codes is not directly dependent on any of the these constructs except the ``proactive awareness'' because one is considered proactively phishing aware if they notice and take into account security indicators of suspicious URLs and resulting websites (e.g. checking before clicking on links or submitting credentials). The SeBIS proactive awareness sub-scale showed good performance and was used for measuring in numerous phishing experiments \cite{Ma, Quinkert, iost2020}. 

We modified the proactive awareness sub-scale questions to focus particularly on behavior around malicious QR codes and created the \textit{Quashing Awareness Scale} (QAS) as shown in Table \ref{tab:QAS}. The first two QAS questions capture the routine behavior when interacting with QR codes analogous to (not) inspecting URLs and visual websites during everyday browsing. The third and forth QAS questions  capture the behavior when faced with security indicators or lack of thereof. Each question is reversely scored and averaged to yield the resultant QAS score. Low scores on the SeBIS score have been related to impulsivity, risk-taking, and dependence (i.e. seeking help from others) \cite{Canfield} and users who did not fail for phish score significantly higher than the rest \cite{Egelman}. 

As phishing and quishing are heavily related, one could reasonably expect the proposed QAS to capture a similar manifestation when measuring quishing awareness as SeBIS, already a validated measure, measures phishing. We also performed a concurrent validity test between the two scales to establish a formal evidence of QAS's validity as part of our pilot study \cite{Egelman2016}. Both scales were significantly correlated  $r(24) =.937, p=0.001$ The QAS scale also showed a reliable internal consistency with Crombach's $\alpha = .728$. 


\begin{table}[hbt]
\renewcommand{\arraystretch}{1.1}
\small
\caption{QAS and SeBIS: Questions}
\label{tab:QAS}
\centering
\begin{tabularx}{\linewidth}{|c|c|X|}
\Xhline{3\arrayrulewidth}
\textbf{\#} & \textbf{Scale} & \textbf{Questions}  \\\Xhline{3\arrayrulewidth}
\multirow{2}{*}{\textbf{1}}  & \textbf{QAS} &  When I scan a QR code, I open the link without first verifying where it goes \\\cline{2-3}
& \textbf{SeBIS} & When someone sends me a link, I open it without first verifying where it goes \\\Xhline{3\arrayrulewidth}

\multirow{2}{*}{\textbf{2}}  & \textbf{QAS} & I know what website I’m visiting based on its look and feel, rather than by looking at the QR code or the URL bar.  \\\cline{2-3}
& \textbf{SeBIS} & I know what website I’m visiting based on its look and feel, rather than by looking at the URL bar.    \\\Xhline{3\arrayrulewidth}

\multirow{2}{*}{\textbf{3}} & \textbf{QAS} & I submit information to QR code websites without verifying it is sent securely (e.g., SSL, “https://”, a lock icon).    \\\cline{2-3}
& \textbf{SeBIS} & I submit information to websites without first verifying that it will be sent securely (e.g., SSL, “https://”, a lock icon)   \\\Xhline{3\arrayrulewidth}

\multirow{2}{*}{\textbf{4}} &  \textbf{QAS} & When scanning QR codes, I read the links to see where they go, before clicking them. \\\cline{2-3}
& \textbf{SeBIS} & When browsing websites, I mouseover links to see where they go, before clicking them. \\\Xhline{3\arrayrulewidth}
\multirow{2}{*}{\textbf{5}} & \textbf{QAS} & If I discover a security problem with the QR code or the website, I continue what I was doing because I assume someone else will fix it. \\\cline{2-3}
& \textbf{SeBIS} & If I discover a security problem, I continue what I was doing because I assume someone else will fix it  \\\Xhline{3\arrayrulewidth} 
\multicolumn{3}{|l|}{\footnotesize  \textit{Scale: Never (1), Rarely (2), Sometimes (3), Often (4), and Always (5).}} \\\Xhline{3\arrayrulewidth}
\end{tabularx}
\end{table}

\subsection{Pilot Quishing}
We conducted a pilot study with 24 participants to verify the pretext, the quishing workflow and protections, and debug the process before recruiting a larger sample. We also used the pilot study to validate the QAS scale. The pilot study parameters were approved by IRB on similar terms: anonymous, standard compensation for 30 minute participation (extended to allow time for the additional questions), consistency/attention checks, and extended debriefing. None of the pilot participants suspected the pretext, the URL, or resulting sign-up website. In the first half of the study, Facebook was the first sign-up option, followed by Google, Twitter, ``new account'', and ``skip.'' The debriefing uncovered that the ``skip'' option enabled participants to immediately avoid using anything related to social media for which they expressed a high skepticism.

We therefore decided to experiment by hiding the ``skip'' option behind an extra click of ``More Login Options.'' This proved a good tactical step from a quisher's point of view since only one user took the pains to click and skip the sign-up. We noticed a lack of interest for the Twitter SSO so we eliminated it in the main study. The pilot study results indicated that Facebook was the most convenient to sign-up with, followed by Google. Participants deemed their Google credentials more ``valuable'' than Facebook stating that they ``\textit{don't mind having their Facebook login shared elsewhere}'' even if it gets compromised. The participants that chose Facebook had the lowest average score $\overline{QAS} = 1.56$ and the ones that skipped the sign-up scored the highest on average $\overline{QAS} = 2.58$.  

\section{Main Quishing Study} \label{section:4}
After the consolidation and consistency checks, we recruited a total of 173 participants. The sample contained 44.2\% female and 55.8\% male participants. Most of them were college graduates (60.1\%), followed by high school graduates (28.3\%), post graduates (10.4\%) and only 2\% participants with less than a high school education. Age-wise, 45.7\% were in the [35-44] bracket, 24.3\% in [25-34], 13.9\% in [45-54], 9.8\% in [55-64], and 6.4\% in [18-24]. Asked about their QR code usage, 32.4\% reported using QR codes ``regularly,'' 56.1\% ``only when QR codes were preffered'' type of information exchange, and 11.6\% said ``only when QR codes were required.''

\subsection{Quishing Susceptibility (RQ1)}
Out of 173 participants, 44.5\% choose Google's SSO for sign-up, 22.5\% choose Facebook, 18.5\% chose to creating a new account, and only 14.5\% choose to skip the registration for the alleged COVID-19 digital vaccination passport trial. The open-ended responses in the survey were coded independently by two researchers on the main decisive factor for their sign-in choice (excellent inter-coder agreement - Cohen's $\kappa = .981$). The Pearson's chi-square test revealed a statistically significant relationship between the sign-up choices and justification factors,  $\chi(3) = 106.978$, $p = .000^{*}$, tabulated in Table \ref{tab:Q1}. Convenience was the factor cited by almost 90\% of the participants that opted for Facebook, quoting that ``\textit{creating an account seemed like a process that might take too long, and I prefer using Facebook to log-in}'' [\textbf{Participant (P)97}]. Interestingly, the participants citing personal safety quoted: ``\textit{I prefer to use Facebook for experimenting with new apps because I don't want spam in my email accounts}'' [\textbf{P39}]. The only participant suspicious of quishing provided an interesting explanation:  ``\textit{I wasn't sure if I can trust this link so I'd rather compromise my Facebook login and then change the password later if needed}'' [\textbf{P17}]. 

\begin{table}[hbt]
\renewcommand{\arraystretch}{1.1}
\small
\caption{Sign-up Choice Justifications}
\label{tab:Q1}
\centering
\begin{tabularx}{\linewidth}{|l|Z|Q|Z|Q|}
\Xhline{3\arrayrulewidth}
\textbf{Factor} & \textbf{Facebook} & \textbf{Google} & \textbf{New Acc.} & \textbf{Skip}  \\\Xhline{3\arrayrulewidth}
Convenience & 35 & 51 & 3 & 7 \\\hline
Perceived Safety & 3 & 10 & 8 & 12 \\\hline
SSO Reluctance & 0 & 0 & 13 & 5 \\\hline
Distrust & 0 & 16 & 4 & 0 \\\hline
Suspicious & 1 & 0 & 4 & 1 \\\Xhline{3\arrayrulewidth}
\textbf{Total} & 39 & 77 & 32 & 25 \\\Xhline{3\arrayrulewidth}
\end{tabularx}
\end{table}

None of the participants that selected Google's SSO was suspicious of quishing, but a good 20\% of them preferred using their Google credentials over Facebook citing: ``\textit{I don’t want to associate this vaccine passport with my Facebook so I chose Google}'' [\textbf{P50}].  Almost 13\% of the participants cited personal safety for their Google choice because they ``\textit{believe Google more than Facebook, to keep [their] data safe}'' [\textbf{P76}].  Anecdotally, one these participants said: ''\textit{I trust the use of my Google account as a passport but also found the site to look reliable enough to be trustworthy. Nothing in the URL seemed out of the ordinary}'' [\textbf{P72}]. The remaining 64\% simply preferred Google because it was the ``\textit{easiest method of logging in}'' [\textbf{P156}].  

Expectantly, the participants that opted for creating a new account predominantly cited a reluctance to use either Google or Facebook for sign-in to ``\textit{link [their] existing accounts with the passport}'' [\textbf{P109}]. Those who opted for personal safety reasons indicated they have ``\textit{privacy concerns}'' about the COVID-19 digital vaccination passport [\textbf{P162}]. Equal percentage of these participants - 12.45\% -  distrusted Facebook and were suspicious of the link/website. A couple of the participants did not trust the website, one questioned the website's visual appearance, and one actually used a browser security indicator to suspecting something is ``phishy:'' 

\begin{description}[leftmargin=!]
\itemsep 0.3em
\item [P37:]``\textit{I don't know how trustworthy the site is. I would prefer to not use my main email in case it is spam.}''

\item [P79:]``\textit{I don't trust that website with my information.  It looks like an official us government site but it's not}''

\item [P123:]``\textit{I’m not familiar with this platform and did not trust it to link to existing accounts I have. Also, the website was not secure in the browser address bar.}''

\item [P166:]``\textit{I don't fully trust the site and didn't want to link my Google account to anything that wasn't 10000\% safe}''
\end{description}

The participants that went the extra step to skip registering  cited time as a convenience in 28\% of the cases, noting that ``\textit{...not creating a login saves me time}'' [\textbf{P163}]. In 48\% of the cases participants cited concerns about their privacy: \textit{I did not want any of my personal accounts to be connected with a vaccine passport because a vaccine passport is private medical information} [\textbf{P21}]. In 25\% of the cases, social media distrust factored the most in skipping the sign-up: ``\textit{I am not comfortable linking social media with my vaccine passport}'' [\textbf{P116}]. Only one participant was suspicious about the COVID-19 digital vaccination passport: ``\textit{I didn't want to log in with my Facebook or Google account because I wasn't sure how safe and secure this website is}'' [\textbf{P113}].


\subsection{Quishing Awareness (RQ2)}
\begin{figure}[!h]
  \centering
  \includegraphics[width=0.8\linewidth]{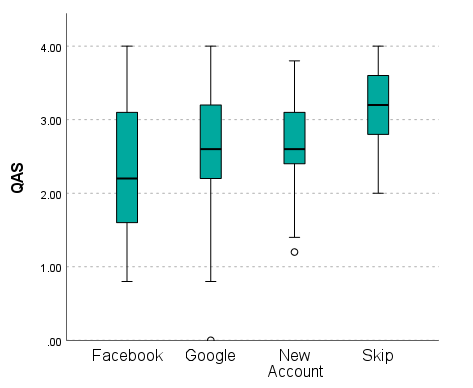}
  \caption{QAS Score Distribution Per Sign-up Choice}
  \label{fig:QAS-means}
\end{figure}

To understand how a possible awareness of threats from malicious URLs associated with QR codes relates to participants' choices for sign-in, we asked the participants to answer each of the QAS scale items from Table \ref{tab:QAS}. Since not all of the choices formally satisfied the test for normality, we performed a non-parametric Kruskal-Wallis test to compare whether there is a difference in the QAS score between the participants from each of the sign-up groups. Indeed, the result was statistically significant $U = 15.335$ \& $p = .002^{*}$ with the distribution of the scores per group shown in Figure \ref{fig:QAS-means}. The participants that chose Facebook scored the lowest on the scale on average($\overline{QAS} = 2.41, \sigma = .895$), followed by the ones that selected Google ($\overline{QAS} = 2.64, \sigma = .765$) and new accounts ($\overline{QAS} = 2.69, \sigma .631$). The highest score on average ($\overline{QAS} = 3.18, \sigma .575$) had the participants that chose to skip the sign-up for the COVID-19 digital vaccination passport. 

When grouped per justification, the suspicious participants scored the highest ($\overline{QAS} = 3.00, \sigma = .551$), followed by the ones that distrusted Facebook ($\overline{QAS} = 2.81, \sigma = .678$), the ones reluctant to use SSO ($\overline{QAS} = 2.72, \sigma = .539$) and the ones concerned about their online safety ($\overline{QAS} = 2.72, \sigma = .647$). The lowest score had the participants that choose convenience ($\overline{QAS} = 2.60, \sigma = .881$) as a reason. Table \ref{tab:Q2} shows the average QAS score per each category of justification for each of the sign-in choices. The participant that chose Facebook but cited lack of trust in the website scored the highest on QAS. So did the participants that chose Google but cited personal safety, although the remaining justifications are on comparable levels of quishing awareness. The participants that chose to create a new account because of convenience or personal safety scored the lowest while the the ones expressing distrust in Facebook or suspicion in the resulting website scored the highest. The participants that chose to skip scored consistently high regardless of the justification provided.

\begin{table}[hbt]
\renewcommand{\arraystretch}{1.1}
\small
\caption{QAS Average Scores and Standard Deviations}
\label{tab:Q2}
\centering
\begin{tabularx}{\linewidth}{|l|Y|Y|Y|Y|}
\Xhline{3\arrayrulewidth}
\textbf{Factor} & \textbf{$\overline{QAS}$} & \textbf{$\sigma$} & \textbf{$\overline{QAS}$} & \textbf{$\sigma$} \\\Xhline{3\arrayrulewidth}
& \multicolumn{2}{c|}{\textbf{Facebook}} & \multicolumn{2}{c|}{\textbf{Google}}\\\Xhline{3\arrayrulewidth}
Convenience & 2.44 & .93 & 2.62 & .83 \\\hline
Perceived Safety & 2.00 & .529 & 2.72 & .518 \\\hline
SSO Reluctance & 0 & 0 & 0 & 0  \\\hline
Distrust & 0 & 0 & 2.72 & .696\\\hline
Suspicious & 2.60 & / & 0 & 0 \\\Xhline{3\arrayrulewidth}
& \multicolumn{2}{c|}{\textbf{New Acc.}} & \multicolumn{2}{c|}{\textbf{Skip}} 
\\\Xhline{3\arrayrulewidth}
Convenience &  2.45 & 1.275 & 3.29 & .62 \\\hline
Perceived Safety & 2.4 & .534  & 3.12 & .617\\\hline
SSO Reluctance & 2.63 & .467 & 3.16 & .572 \\\hline
Distrust & 3.25 & .678 & 0 & 0\\\hline
Suspicious & 3.05 & .66 & 3.20 & /\\\Xhline{3\arrayrulewidth}
\end{tabularx}
\end{table}

\begin{figure}[!h]
  \centering
  \includegraphics[width=0.8\linewidth]{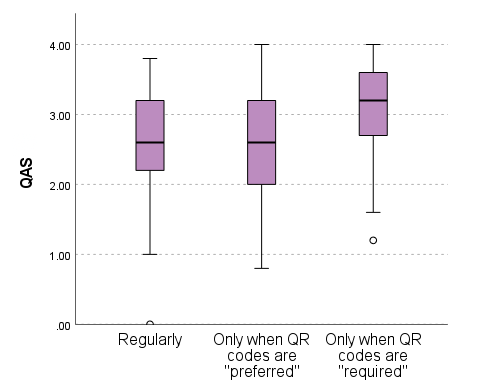}
  \caption{QAS Score Distribution Per QR Code Usage}
  \label{fig:QAS-usage}
\end{figure}

We also performed a Kruskal-Wallis test to compare whether there is a difference in the QAS score between the categories of QR code usage or participants' demographic identities. The QAS score has a statistically significant difference only between the categories of QR usage as shown in Figure \ref{fig:QAS-usage}, $U = 6.594$ \& $p = .037^{*}$. Participants that used QR codes regularly had the lowest average score ($\overline{QAS} = 2.60,  \sigma = 0.779$) but not far away were the participants that opted for using QR codes as a preferred touch-free information transfer ($\overline{QAS} = 2.64, \sigma = 0.762$). The participants that used QR codes only when required scored noticeably higher on average ($\overline{QAS} = 3.08, \sigma = 0.774$). Participants' level of education, gender identity, or age did not significantly factor in participants' quishing awareness scores.  

\subsection{Quishing Implications}
Our first research question asked for the susceptibility factors that contribute in a plausible quishing campaign for harvesting credentials. We did not harvest actual credentials as that constitutes more than a minimal harm to our participants, but we were able to determine the factors driving participants' willingness to yield credentials to a quishing website, nonetheless. Convenience, perhaps unsurprisingly, is the main factor contributing in successful quishing as 55.5\% of the participants took the route of least resistance for transfer of credentials. The perception of online safety seems a promising factor that helped 11.5\% of the participants to avoid yielding their Facebook/Google credentials, but we uncovered a worrisome trend of differential perception between Facebook (less safe) and Google (more safe) in 7.5\% of the cases that still renders users susceptible to quishing. 

Another ameliorating factor against quishing could be the reluctance to use Facebook/Google credentials by another 7.5\% of the participants that opted to create a new account, but users tend to reuse credentials so quishers might still be able to harvest useful information \cite{Pearman}. Distrusting social media again could wield some anti-quishing leverage for users, but 4/5 of the 11.5\% of participants that cited this factor chose to use their Google credentials. The most worrisome finding of our study, at least in the nascent stages of quishing, is that only 3.4\% of the participants were suspicious of deception for credential harvesting. 

It appears that crafting an authoritative pretext is still a sufficiently potent persuasion factor, regardless of the shift in phishing modality from an email to a QR code \cite{Krombholz2015}. Quishing, similarly as its older sibling phishing, appears to thrive in conditions of scarcity as the requirements for proof-of-vaccination proved influential in increasing the credibility of the quishing pretext \cite{Parsons}. The social proof and reciprocation were perhaps less salient persuasion factors, but we did not find any evidence explicitly countering their effect. One could argue that the considerable COVID-19 anti-vaccination sentiment \cite{Pullan} could render all persuasion factors unusable in the particular quishing context and that certainly is a line of future inquiry we want to pursue. 

The lack of quishing awareness also worked to the advantage of the quishers in our study, as we also employed the QAS scale to answer the second research question. With an overall average score across the sample of $\overline{QAS} = 2.68, \sigma = 0.78$, only the participants that selected to skip the sign-up citing the convenience factor had scored higher on average. We saw the same outcome when considering the perceived safety as a factor with an exception of a $+.04$ in the average score of the participants that selected to sign-in with their Google credentials. The QAS scores associated with the reluctance of using Google/Facebook credentials and distrust in Facebook are on par or higher with the sample average, revealing a positive awareness of phishing and online deception in general. The suspicious participants, except the participant choosing Facebook, scored $.37$ and $.52$ higher than the average when choosing a new account or skipping registration, respectively. 

Quishing awareness is related to the frequency one uses QR codes as revealed by our analysis. As show in Figure \ref{fig:QAS-usage}, participants that used QR codes regularly or preferred them over other options for information transfer scored lower on the QAS scale compared to the participants that used QR codes only when required. The latter ones, however, constituted only 11.6\% of our sample. Interestingly, only 15\% of the participants chose to skip the registration, yielding only 1.7\% of the sample entirely insulated from quishing by avoiding using QR codes and transferring credentials online. These findings reveal that quishing is a relevant threat given that users embrace or have no objection to the adaptation of QR code-based interaction imposed by the COVID-19 pandemic. Whether these results will hold in future certainly depends on how the adoption of QR codes will unfold and we are equally interested in exploring any induced shift in quishing.   

\section{Anti-Quishing}  \label{section:5}
Changing users' behavior around potentially malicious QR codes was the imperative of our third and forth research questions, considering the surge in quishing attacks. As is the case with anti-phishing, the two logical approaches would be to: (1) train users how to spot a quishing campaign; and (2) develop security indicators within the QR code interaction elements to help users avoid falling for quish \cite{Wash}. Phishing training is usually delivered through facts-and-advice materials \cite{Wash}, gamification or simulation for malicious URL detection \cite{Sheng, Yang, Canova} or embedded phishing exercises \cite{Siadati}. We certainly could not develop them all at once; instead, we contextualized each training modality to fit in a quishing setting in subsection \ref{subsection:5.1}. More practically, we propose several designs for quishing-appropriate security indicators in subsection \ref{subsection:5.2} utilizing the previous work in security cues, warnings, and anti-phishing nudges \cite{Porter-Felt, Redmiles, Vance, Reeder}.

\subsection{Quishing Awareness Training (RQ3)} \label{subsection:5.1}
The phishing training, in each of the aforementioned modalities, assumes at least a minimum shared email/browsing experience from the trainees to contrast legitimate and malicious emails/URLs. A similar assumption might not entirely hold for a quishing training, given that the QR codes apply in a wider range of information transfer that many ``trainees'' might have not yet encountered. Take for example the most recent release of the Coinbase QR Code commercial at the Super Bowl LVI \cite{Coinbase}. For most of the 60 second commercial block, users saw a black background with a QR code (in multiple monochrome colors) bouncing from corner to corner reminiscence of old screensavers (in the last moment, the commercial invited viewers to ``Get \$15 in free Bitcoin for signing up. Plus, a chance to win \$3 million prizes!''). A deliberate choice to turn a legitimate QR code suspicious probably makes marketing sense by working on viewers' curiosity and reciprocity, but creates a confusing experience for users accustomed to paper QR codes for menus or electronic QR codes for movie tickets.  

The facts-and-advice ``stories'' should therefore not just convey lessons but also cover quishing pretexts in detail. The lessons themselves are similar to the ones for phishing \cite{Wash} and be logically adopted as a foundation in a basic quishing training. We propose one such adaptation in Table \ref{tab:Lessons}. The lessons fit well with our COVID-19 digital passport trial and could equally materialize in a pretext where a victim is sent a legitimate link to a video resembling the commercial above where the QR code leads to an attacker-controlled cryptocurrency website with a more lucrative deal (usually clustered with cryptocurrency scams \cite{Phillips}).    

\begin{table}[hbt]
\renewcommand{\arraystretch}{1.1}
\small
\caption{Basic Quishing Lessons}
\label{tab:Lessons}
\centering
\begin{tabularx}{\linewidth}{|c|X|}
\Xhline{3\arrayrulewidth}
\textbf{\#} & \textbf{Lesson}  \\\Xhline{3\arrayrulewidth}
\textbf{1} & Malicious QR codes can come in both physical and digital form; They could resemble QR codes from restaurants, posters, or parking meters, but could also be attached in IT department emails, shown on websites and social media platforms, or even TV commercials
\\\hline

\textbf{2} & Read the link in the notification after scanning a QR code to see where it really goes to
\\\hline

\textbf{3} & If you click a link right after scanning a QR code, your identity can be stolen
\\\hline

\textbf{4} & If you click a link after scanning a QR code, make sure to look for security indicators preceding the URL in the browser like `Not Secure' tags, exclamation marks, strike-through words or padlock icons, or red warning screens
\\\hline

\textbf{5} & Malicious QR codes enable `quishing' or phishing with QR codes. Phishing is when an attacker sends you a fake email; Quishing is when an attacker gets you scan a fake QR code \\\hline

\textbf{6} & Quishing is your problem because if you click on the link after scanning a fake QR code, it is your information being stolen \\\Xhline{3\arrayrulewidth}
\end{tabularx}
\end{table}

The myriad of quishing pretexts and the nature of QR code interactions make a more convincing case for a training delivered through gamification. Traditional phishing training is usually delivered online (e.g. a gaming website or an app). QR codes allow for extending the game in physical spaces, akin to alternate reality games (ARGs) \cite{Flushman}. QR codes could easily enable application of the \textit{reflection} game design principle \cite{Canova} where ``players'' stop and reflect after encountering and scanning such a code, e.g. a QR code sticker over a regular billboard or a QR code icon bouncing from corner to corner on a digital billboard at a bus stop. QR codes also fit with the \textit{story-based agent environment} principle - a ``player'' doesn't have to be a ``fish'' but could be a ``papparazzi'' agent scanning QR codes around exclusive places, for example. Or become a ``forensic investigator'' agent looking for both physical and URL tampering of QR codes, as in other security-related ARGs \cite{Morelock}. Since quishing is a form of phishing, the \textit{conceptual–procedural} principle is easy to satisfy with facts-and-advice procedural phishing knowledge and iteratively build the concept of malicious QR codes. 

ARGs are versatile enough to extend the embedded phishing exercises usually applied in working environments \cite{Siadati}. Employees could be periodically ``quished'' with emails containing malicious codes as in \cite{Chouinard}, but an ARG could help employees' retain and expand their general phishing knowledge. An ARG with posters including training QR codes in common rooms, company's parking garages, or even collaborative platforms like Slack could increase both quishing and phishing awareness over a longer period of time, provided that measures based interactive examples perform best in reminding employees about online deception \cite{Reinheimer}. QR codes are low-cost extension of the general phishing awareness training and companies could also capitalize on their internal adaptations to accommodate QR codes since the COVID-19 pandemic start.  

\subsection{Quishing Security Indicators (RQ4)} \label{subsection:5.2}
In our study, only one participant out of 173 - \textbf{[P123]} - noticed the security indicator preceding the \texttt{covid-passport.us/qr} URL. Evidence suggests that users' often find security indicators confusing and irrelevant \cite{Reeder}. Users also ``habituate'' to repetitive exposure or security warnings online \cite{Vance}. To rectify this problem, security indicators deliberately include ``design frictions'' designed to disrupt automatic interactions, and grab the attention of the users \cite{Cox}. Users first need to need to pay attention with their camera to lock on the QR code and next on the screen to capture the temporal notification to be able to open the embedded URL. Since users' need to take explicit action before the URL is displayed in a browser, a security indicator about the deceptive nature of the URL immediately before would grab their attention. Users do heed forced attention browser warnings \cite{Petelka}, so we believe this could also be the case when scanning QR codes too.  

\begin{figure*}[tb]
  \centering
  \includegraphics[width=\linewidth]{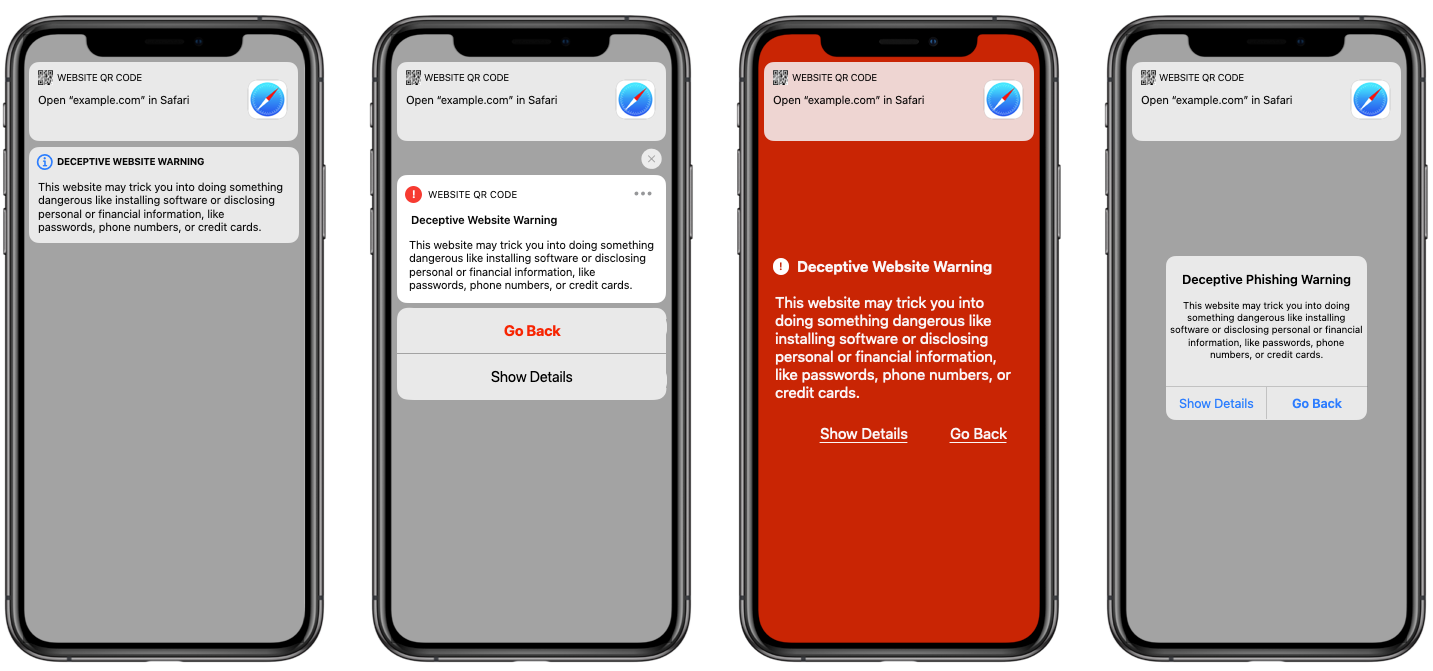}
  \caption{Quishing Security Awareness Indicators: (a) additional notification; (b) extended notification with actions; (c) background screen with actions; (d) alert with actions}
  \label{fig:QAS-indicate}
\end{figure*}

Following this approach, we developed four possible security indicators to grab users' attention when scanning QR codes, shown in Figure \ref{fig:QAS-indicate}. We did use iOS as a basis, but the overarching idea of each of the frictions could be easily translated in Android or other mobile operating systems. All four security indicators use the standard Safari browser warning heading ``Deceptive Website Warning'' and follow up text regularly shown when users encounter suspected websites in Safari \cite{Safari}. We use an visual/action-based inhibitors in variable degrees during the user QR code interaction to meet a wider range of user preferences for security warnings ~\cite{Wogalter}.  

The first security indicator (A) is a simple notification after the standard notification users receive when scanning a QR code. Here, we employ minimal yet meaningful visual friction with an indicator that contains a blue information icon and blends with the notification structure in iOS. The second security indicator (B) increases the visual friction with the red exclamation icon attached to application tiles in iOS and offers the possibility for a user to take either a ``Go Back'' or ``Show Details'' actions. The third security indicator (C) maximizes the visual friction by displaying a standard red browser warning in the background over the entire screen with the same two action options. The last security indicator instead uses the alert popup in iOS as an interaction friction to warn the user and explicitly demand either of the actions to be taken before the user proceeds.

We conducted a follow-up usability study with another sample of 124 participants where we tested each of the security indicator options. This study was also approved by the IRB of our institution. We recruited Amazon Mechanical Turk participants that were 18 years or above old, with iPhone as a primary mobile device, and have scanned QR codes with their iPhone to open URLs. The study was anonymous and took around 15 minutes to complete with the standard compensation rate. We first collected data to calculate their QAS scores, then exposed them to the concept of quishing and collected their preferences (up to two). The sample contained 44.4\% female and 55.6\% male participants. Most of them were college graduates (58.1\%), followed by high school graduates (32.3\%), post graduates (8.9\%) and only 0.1\% participants with less than a high school education. Age-wise, 44.4\% were in the [35-44] bracket, 24.2\% in [25-34], 12.9\% in [45-54], 15.3\% in [55-64], and 3.2\% in [18-24].

Table \ref{tab:SecInd} shows the distribution of the user preferences with the average QAS score and standard deviation for each choice. The security indicator (C) was the most preferred one by the participants that, interestingly, scored lower than the average QAS of the sample. The attention grabbing effect of the red screen was obvious and useful to participants: ``\textit{It grabs your attention. Little boxes that look like every other notification often go unread and get clicked on haphazardly}'' [\textbf{P61}]. Participants also preferred the security indicator (C) together with the security indicator (B), scoring a bit higher than the previous group, but still bellow the average QAS of the sample. Here, participants suggested what actually they would like to see as a security warning: ``\textit{Option B was my favorite because it offered me details. Option C I like because the red screen was very obviously something bad but it wasn't my first choice because I feel like it could be overly worrying for some people}'' [\textbf{P35}]

The third most preferred security indicator was (B) and this group of participants scored the lowest of all groups on the QAS scale. Participants like this option because: ``\textit{It incorporates red, which I quickly associate with a problem. It doesn't crowd the screen. It has two action buttons easily accessible}'' [\textbf{P7}]. The group of participants the scored the highest on the QAS scale preferred both the security indicator (B) and (D). They praised the balancing effort of visual frictions with explicit actions:  \textit{I think they're the most concise, and they look the most modern and less `scary' than the other options. They seem the most normal, they feel familiar.}'' [\textbf{P43}]. 

Participants preferred the security indicator (D) alone because: ``\textit{It provided enough details but also gave me a choice on how to continue}'' [\textbf{P123}], or together with (C) because: ``\textit{Those two are much more emphatic in letting me know to be suspicious. If a user is distracted or just not thinking 100\% through their process, the other warnings might not be emphatic enough to grab their attention} [\textbf{P121}]. The participant that chose the security indicator (D) together with (A) disliked the red color and the one preferring the security indicator (A) stated: ``\textit{I would rather make the choice myself on what to do instead of being blocked completely}'' [\textbf{P2}].

\begin{table}[hbt]
\renewcommand{\arraystretch}{1.1}
\small
\caption{Security Indicators Preferences}
\label{tab:SecInd}
\centering
\begin{tabularx}{\linewidth}{|l|Y|Y|Y|}
\Xhline{3\arrayrulewidth}
\textbf{Preference} & \textbf{Count} & \textbf{$\overline{QAS}$} & \textbf{$\sigma$} \\\Xhline{3\arrayrulewidth}
A	& 1	& 2.2 & / \\\hline
A and D	& 1	& 2.4 & / \\\hline
B	& 17 &	2.33 & .647 \\\hline
B and C &	21	& 2.52	& .601\\\hline
B and D	& 5	& 2.92 & .228 \\\hline
C	& 74 &	2.39 & 523\\\hline
C and D	& 3 &	2.60 & .6\\\hline
D	& 2	& 2.50	& .424 \\\Xhline{3\arrayrulewidth}
\textbf{Total} & \textbf{124} & \textbf{2.6}  &  \textbf{.584} \\\Xhline{3\arrayrulewidth}
\end{tabularx}
\end{table}

Only nine participants confirmed they encountered suspicious QR codes in the past with the second highest quishing awareness score ($\overline{QAS} = 2.80, \sigma = .435$). Three of them cited they received QR codes via an instant message requesting bill payments to be submitted to the embedded URL, couple of them pointed to QR codes associated with cryptocurrencies, and the rest refereed to physical QR codes. Four of them explicitly stated they suspected the resulting website because of a security indicator associated with it: ``\textit{I clicked the QR code link that took me to a `spammy' looking website that resulted in security warnings on my phone.}'' [\textbf{P93}]. 

Controlling for QR code pattern of usage, 14.4\% indicated they use QR codes regularly and preferred the security indicator (C) the most. They scored the lowest on average on the QAS scale of all groups: $\overline{QAS} = 2.22, \sigma = .508$. The 16\% of the participants that preferred touch-free information transfer weren't much better on the QAS scale, scoring $\overline{QAS} = 2.29, \sigma = .412$, but they were roughly evenly split between the preference for the security indicator (C) and (B). The remaining 59.6\% of the participants that used QR codes only when required scored the highest between the three usage groups on average: $\overline{QAS} = 2.52, \sigma = .568$. The preferences of this group were similarly distributed as in Table \ref{tab:SecInd}.

Controlling for gender identity, we noticed that the female participants preferred the security indicator (B) alone, and together with (D), more than the male participants ($\overline{QAS} = 2.40, \sigma = .494$). The male participants were overwhelmingly in support of the individual security indicator (C) and together with the other three options ($\overline{QAS} = 2.47, \sigma = .589$). Respective to the level of education, we noticed that the post-graduate degree participants overwhelmingly preferred the security indicator (C), while the other participants were more balanced in their preferences. Interestingly, they scored the lowest on QAS scale  ($\overline{QAS} = 2.35, \sigma = .629$). Age-wise, the older the participants were, the more the stronger the preference of the individual indicator (C).  Similarly, the QAS score decreased with the age, with the [18-24] bracket the highest ($\overline{QAS} = 2.85, \sigma = .1$).






\section{Discussion}  \label{section:6}
Our quishing study uncovered important field evidence about this emerging type of phishing. The level of quishing susceptibility that we observed is on par or higher than the one for phishing, but this is perhaps to be expected given the early days of mass quishing \cite{Greitzer}. People's vigilance, or the ability to detect anomalies for a sustained period, has been poor when it comes to phishing, so it is hardly probable for a positive change to happen when dealing with malicious QR codes \cite{Canfield2016, Wash2020}. This is inline with the evidence that the majority of our participants readily trumped convenience over security. Even more, almost 60\% of the participants in the follow-up study chose security indicator (C) which visually stands out the most, reasoning that a stronger friction is needed to turn their attention towards a potential anomaly with the embedded URL \cite{Bravo-Lillo}. These two groups scored lower-than-average on the quishing awareness scale, suggesting that QR code are seldom associated with anomalies or even negative consequences of clicking the embedded URL \cite{Greitzer}. 


Our findings also points out a small but important set of suspicious users showing a proto-vigilance to quishing. Scoring among the highest on the QAS scale in both the main and the follow up study, these participants modeled their QR code behavior from what they know about or previously experienced with phishing, but rarely utilized security indicators. This probably results more from \textit{neuroticism} or the anxious resistance to sharing information rather than from \textit{conscientiousness} or attention to detail, the two personality traits associated with resistance to phishing \cite{Workman}.

Our follow-up usability study confirms that quishing works because the security indicators offer a poor cost/benefit trade-off to users \cite{Sunshine}. First, a minute spent examining an URLs results in an estimated cost (in terms of user time) of two orders of magnitude greater than all phishing losses \cite{Herley}. Second, the cost of learning the meaning of a security indicator outweighs the benefit of seamless interaction \cite{Reeder}, Third, the cost of phishing losses, e.g. compromised credentials, seems to decrease, as few of participants put it: ``\textit{I don't mind having my Facebook login compromised, I can always reset the password.}'' The time spent on scanning a QR code, on the other hand, is an acceptable cost and perhaps the security indicators should factor the temporal trade-off when warning about quishing ~\cite{Krombholz2015}.

\subsection{Limitations}
We note several limitations of our study. Repeating the study with a larger sample will add more evidence not just about the evolution of the quishing pretexts and tactics, but also provide further validation of the QAS scale. We used a COVID-19 pretext but future studies could uncover variable degrees of susceptibility to quishing moderated by new pretexts and overall increase in quishing awareness. We did not harvested actual credentials and it is plausible that many of the participants would have avoided the sign-up when encountering the impersonated Facebook and Google websites. We used iOS for quishing security indicators and users of other mobile operating systems might express preferences that do not conform with the trend observed in our follow-up study. It is equally plausible that new and enhanced quishing indicators will be developed in future that would render the ones proposed in this study obsolete. 

Our participants were 18 years or older and from the US. However, smartphones are prevalent among teenagers \cite{Ionut} and they also encounter QR codes on increased basis (e.g. scavenger hunts \cite{Gressick}). Our results might not entirely generalize for age groups below 18 years given their specific smartphone usage patterns \cite{Adelhardt}, despite the fact that teenagers show poor performance in detecting phishing \cite{Nicholson2020}. Similar limitation holds for users' country of residence as phishing susceptibility differs from country to country \cite{Gopavaram}. 

We used simple QR codes in our study, however, the beautification of QR codes is an active area of development where images, dimensions, or reflection effects are incorporated to further personalize the layouts \cite{Xu}. All of these factors could affect how a user approaches to scan a QR code in a first place and access the embedded URL for potential quishing. We also performed the study when no considerable proposal for creating a secure QR code ecosystem exists. Novel security primitives for QR codes and regularly updated lists of quishing URLs and domains for automated quishing detection could make future quishing difficult ~\cite{Acharya}.

\subsection{Ethical Considerations}
Every public phishing study runs the risk of informing the real-world phishers about the actionable anti-phishing plans. Our study is no exception and we are aware that the preliminary results suggesting high level of susceptibility and relatively low quishing awareness could amplify the quishing attacks in the wild. We believe, however, that the benefits of this field study significantly outweigh the potential quishing harm, especially our actionable proposals for anti-phishing training and security indicators. One could object that the COVID-19 vaccination pretext might have negative consequences in dissuading users against getting a vaccine and thus undermine the collective mass immunization effort. We did not observe such an effect and we could not control how each reader internalizes our findings. Phishers already employ COVID-19 as a pretext at large  for some time \cite{CDC-Phishing}, so our study should not tip the vaccination scales considerably.

\section{Conclusion}  \label{section:7}
Quishing is as much of a real issue as its big brother phishing. As in phishing, individuals are willing to compromise credentials for the sake of convenience when accessing malicious URL embedded in QR codes. Individuals do lack quishing awareness, but unlike phishing, neither training nor advice is readily available yet. We took the opportunity to develop a scale to measure the quishing awareness and use the results to create phishing training guidelines and test quishing security indicators. We hope that our results will help the security community towards an actionable anti-quishing effort that incorporates some of the suggested anti-quishing efforts.

\bibliographystyle{ACM-Reference-Format}
\bibliography{\jobname}


\end{document}